\documentclass[prl,aps,floats,superscriptaddress,floatfix,twocolumn]{revtex4}
\usepackage{amssymb,amsmath}
\usepackage{amsmath,amssymb}
\usepackage{graphicx}
\usepackage{psfrag}
\usepackage{color}
\usepackage{soul}
\usepackage{dcolumn}
\usepackage{mathtools}

\usepackage{bm}
\usepackage[normalem]{ulem}

\def\beq{\begin{equation}}
\def\eeq{\end{equation}}
\def\bea{\begin{eqnarray}}
\def\eea{\end{eqnarray}}
\begin{document}
 \title{
 Quenched activity induces nonuniversal scaling in  nonreciprocal XY Models and 
surfaces}
 \author{Sudip Mukherjee}\email{sudip.bat@gmail.com}
\affiliation{Barasat Government College,
10, KNC Road, Gupta Colony, Barasat, Kolkata 700124,
West Bengal, India}
\author{Abhik Basu}\email{abhik.123@gmail.com,abhik.basu@saha.ac.in}
\affiliation{Theoretical Physics Division, Saha Institute of
Nuclear Physics, 1/AF Bidhannagar, Calcutta 700064, West Bengal, India}

\begin{abstract}
Active XY models and active surfaces are two paradigmatic nonequilibrium systems with distinct microscopic origins. We show that the hydrodynamic theories for a quenched-disordered nonreciprocal random bond two-dimensional XY model and an  inversion-symmetric active surface tangentially advected by quenched velocities are identical. This theory predicts sub-logarithmic phase order in the XY model and sub- or super-logarithmic positional order in the surface for short-range disorder with nonuniversal exponents, which vary continuously with the degree of transversality of the disorder variance. 
We argue that the nonreciprocal random bond XY model can disorder through vortex proliferation.

\end{abstract}
\maketitle

The two-dimensional (2D) equilibrium XY model is exceptional, because in its low-temperature ordered phase with quasi long-range order (QLRO), the spin correlation function is characterized by a model parameter dependent exponent~\cite{KT,chaikin}. 
Studies on a random quenched bond disordered generalization of the 2D XY model suggest that such short ranged bond disorders are largely irrelevant perturbations to the QLRO in the 2D XY model; see, e.g. Refs.~\cite{deng,raja}. Nonreciprocity, which seemingly violate the action-reaction symmetry, are hallmark of nonequilibrium systems. How such interactions affect nonequilibrium
steady states and phase transitions form a rapidly growing
research area~\cite{nonrec1,nonrec2,nonrec3,nonrec4} at present. Recent studies on 2D nonreciprocal XY models  reveal  possibilities of long-range order (LRO)~\cite{klapp}, presence of asters and shock lines in the steady states~\cite{solon}, destruction of polarized states by defects~\cite{sriram-defect}, pinning of the spin orientation along preferred lattice directions due to anisotropy~\cite{gambassi}, nonreciprocal effects on defect annihilation~\cite{levis} and nonreciprocal frustrations giving order~\cite{hanai}. Active surfaces, another paradigm for 2D active systems, are thin, deformable interfaces 
that continuously consume energy to maintain non-equilibrium behaviour~\cite{bassereau-book} including instabilities and super stiffness ~\cite{tirtha-njp} that have no equilibrium counterparts~\cite{nelson-book}. 

In this Letter, motivated by a broad range of systems, e.g., disordered arrays of synthetic microrotors ~\cite{micro-rot} and vibrating spinners~\cite{spinner}, arrays of nonreciprocal Josephson junctions~\cite{joseph1,joseph2} and nonreciprocal active metamaterials~\cite{meta-mate,meta-mate2}, we construct a generic 2D nonreciprocal, isotropic random bond XY model (hereafter NRB XY model) on a substrate. If $J_{ij}$ is the bond interaction on the spin at site $i$ due to its nearest-neighbor spin at site $j$, then $J_{ji}=-J_{ij}$ is the corresponding interaction on the spin at $j$ due to the spin at  $i$.
%
Such ``nonreciprocal'' interactions ensure that the spin fluctuations do not relax towards the minimum of a global energy function, or equivalently, detailed balance is broken. We derive the hydrodynamic equation for the local phase $\theta({\bf x},t)$ in this NRB XY model, which 
 is also the hydrodynamic equation for conformation (height) of a fluctuating active surface tangentially advected by  random quenched velocities, a minimal theoretical description with relevance to a broad class of fluctuating surfaces, including lipid membranes with permanently cross-linked BAR-domain proteins~\cite{bassereau1} or immobilized transmembrane protein complexes~\cite{madsen}, bilayers attached to a static anisotropic cytoskeletal or substrate network~\cite{sackmann}, supported lipid bilayers on heterogeneous substrates~\cite{watkins}, and growth through porous media~\cite{krug}.
 

 In the absence of noise $\theta({\bf x},t)=const.$. However, in the presence of noise $\theta({\bf x},t)$ is not a constant any more, rather $\theta({\bf x},t)$ is a slowly varying function of $\bf x$ and $t$. The hydrodynamic equation that we derive shows that the fluctuations of $\theta$ diverge with the spatial extent $L$ of the system according to the law
\begin{equation}
 \langle\theta({\bf x},t)^2\rangle = A\bigg[\ln\bigg(\frac{L}{a_0}\bigg)\bigg]^\eta, \label{basic}
\end{equation}
where the amplitude $A$, microscopic cutoff $a_0$ and most importantly, the exponent $\eta$ are all {\em nonuniversal} parameters of the system. Although nonuniversal, $\eta$ has a lower bound: $\eta\geq 1/2$. The fact that $\langle\theta({\bf x},t)^2\rangle$ does diverge for $L\rightarrow \infty$ for all allowed values of $\eta$ simply means that 2D NRB XY model does not have true orientational LRO, but can be infinitely more ordered ($\eta<1$) than the 2D XY model, or short-range ordered for $\eta>1$. Similarly, an active surface described by this equation can be infinitely more  ($\eta<1$) or less ($\eta>1$) positionally ordered than a fluctuating Edward-Wilkinson (EW) surface~\cite{ew,stanley}. Notice the difference in the scaling properties of the two realizations of the model for $\eta>1$, which is due to the fundamental difference in the nature of $\theta$ - compact in the NRB XY model versus noncompact in the surface.

\begin{figure}[htb]
 \includegraphics[width=0.51\columnwidth]{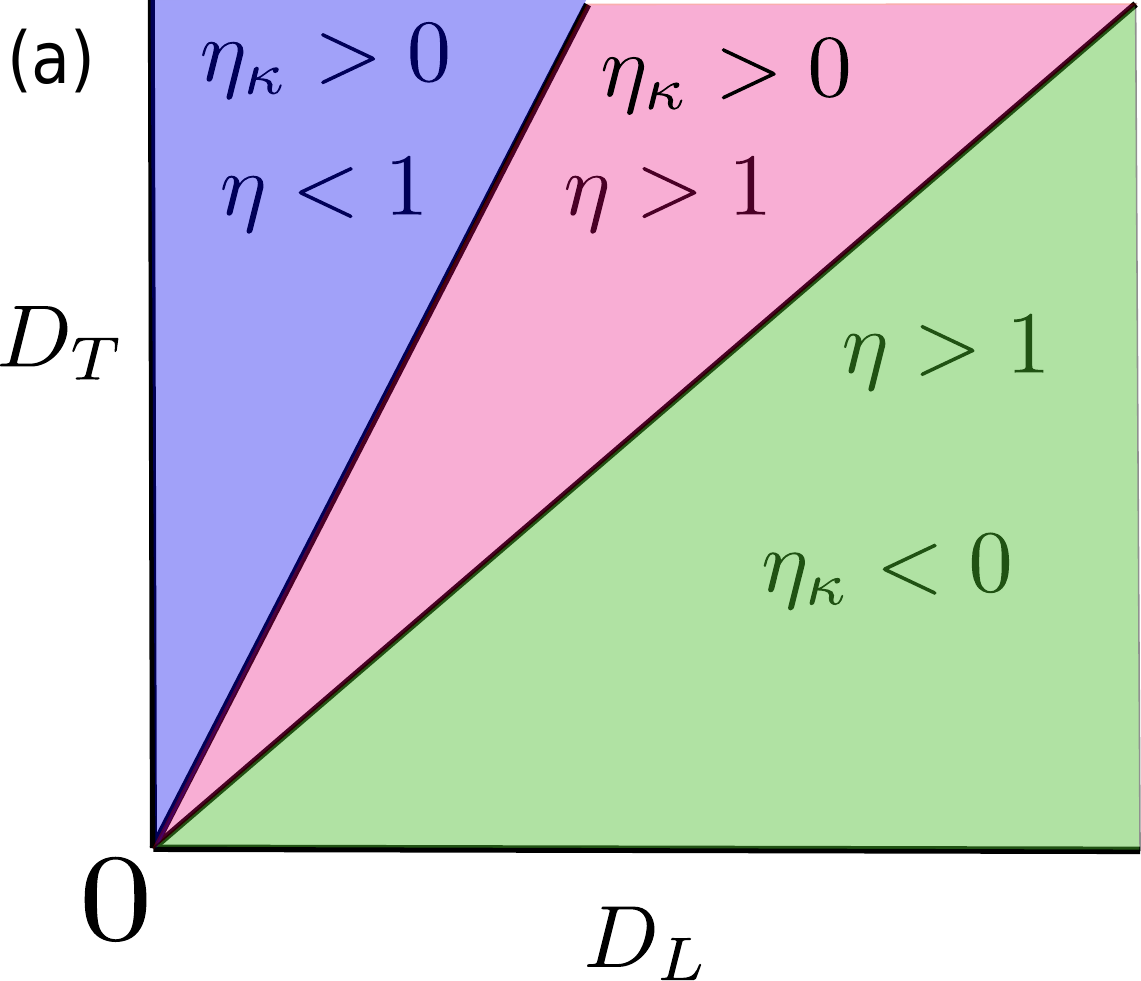}\hfill\includegraphics[width=0.47\columnwidth]{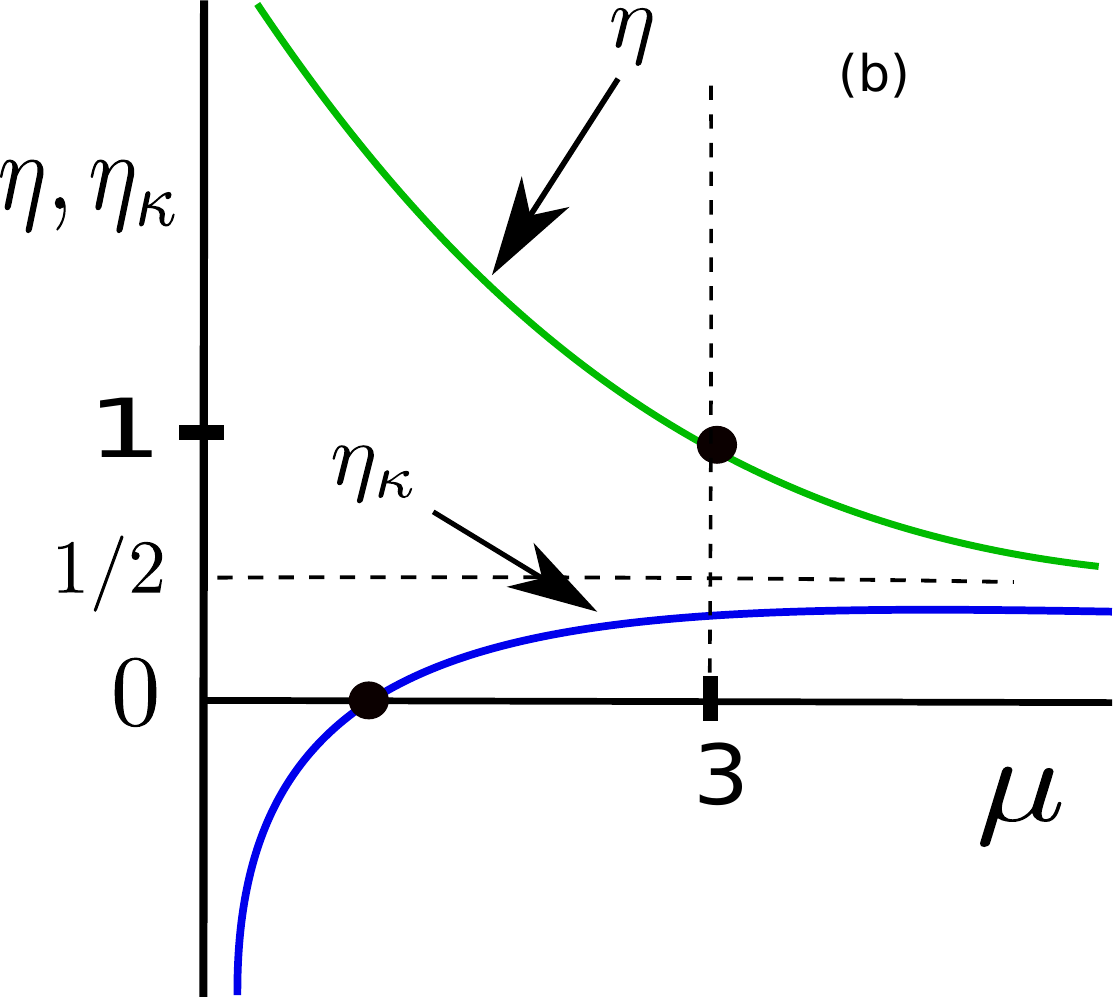}
 \caption{(left)Phase diagram of the model in the $D_L-D_T$ plane. The blue region with $\eta<1,\eta_\kappa>0$ has strong quasi-ordered phase and logarithmically faster than diffusive dynamics, the pink region with $\eta>1,\eta_\kappa>0$ has weak quasi-ordered phase and logarithmically faster than diffusive dynamics, and the green region with $\eta<1,\eta_\kappa<0$ has weak quasi-ordered phase and logarithmically slower than diffusive dynamics. We conjecture that the phases in the  pink and green regions will be destroyed by vortex proliferation in the 2D NRB XY model. See text.  (right) Variation of the exponents $\eta$ (green) $\eta_\kappa$ (blue) with $\mu$. Both $\eta,\eta_\kappa$ approach 1/2 for $\mu\rightarrow\infty$ (horizontal broken line). Black dots represent $\eta=1$ (QLRO) for $\mu=3$ and $\eta_\kappa=0$ (pure diffusive dynamics)   for $\mu=1$, highlighting that these two are  two different points in the parameter space. If our conjecture on vortex proliferation for $\eta >1$ (i.e., $\mu>3$) is correct, then only the parts of the $\eta$- and $\eta_\kappa$-curves falling on the right of the vertical broken line will be observed in NRB XY model. See text.  }\label{phase}.
\end{figure}

We conjecture, and support below with heuristic arguments, that vortices in NRB XY models are irrelevant, at least for sufficiently weak noise, when $\eta<1$, i.e., in NRB XY models exhibiting stronger order than 2D equilibrium XY models. The same arguments imply that vortices remain relevant, even at arbitrarily weak noise, when ($\eta>1$). This argument identifies $\eta = 1$ as the phase boundary separating relatively well-ordered states, whose phase fluctuations are described by \eqref{basic}, from much less ordered states, which are presumably characterized by short-range order and unbound vortices. Consequently, in the 2D NRB XY model, only a finite range of $\eta$ values is expected to be physically realizable in practice:
%
\begin{equation}
 \frac{1}{2}\leq \eta\leq 1. \label{basic-xy}
\end{equation}
In a surface, however, there are no vortices. As a result, it remains stable even for $\eta>1$, giving less positionally ordered phases than an EW surface. We call this {\em weakly ordered states}. The $\eta<1$-region, also accessible by a stable surface gives {\em strongly  ordered states}. 
If our conjecture that regimes with $\eta>1$ in the NRB XY model are unstable to vortex proliferation is incorrect, then the regions with $\eta>1$ would instead exhibit weaker order than QLRO.

In the strongly phase-ordered states for both 2D NRB XY model and the active surface with quenched tangential velocities with $\eta<1$, fluctuations relax {\em logarithimcally} faster, characterized by another nonuniversal exponent $\eta_\kappa$, than ordinary diffusion. The weakly ordered states with $\eta>1$, accessible to surfaces and if also to NRB XY model, can have both faster or slower relaxation of fluctuations depending upon the location in the parameter space. 

Our theory also predicts the phase diagram shown in Fig.~\ref{phase},
where the two axes are the two non-negative parameters with a dimensionless ratio $\mu$ that define the disorder distribution as defined more precisely below. 
The topology of the phase diagrams are expected to be same as those found in simulations or possible experimental realizations of the model, assuming that the parameters of our hydrodynamic theory are smooth, continuous and single-valued functions of the parameters in simulations and controlled experiments.

We speculate that as $\eta (\mu)$ passes through $\eta=1$, a vortex unbinding transition takes place. Making reasonable assumptions, which are discussed later, we find that the correlation length diverges extremely strongly as the transition is approached at $\eta(\mu)=1$ for $\mu=3$.

\begin{figure}[htb]
 \includegraphics[width=0.45\columnwidth]{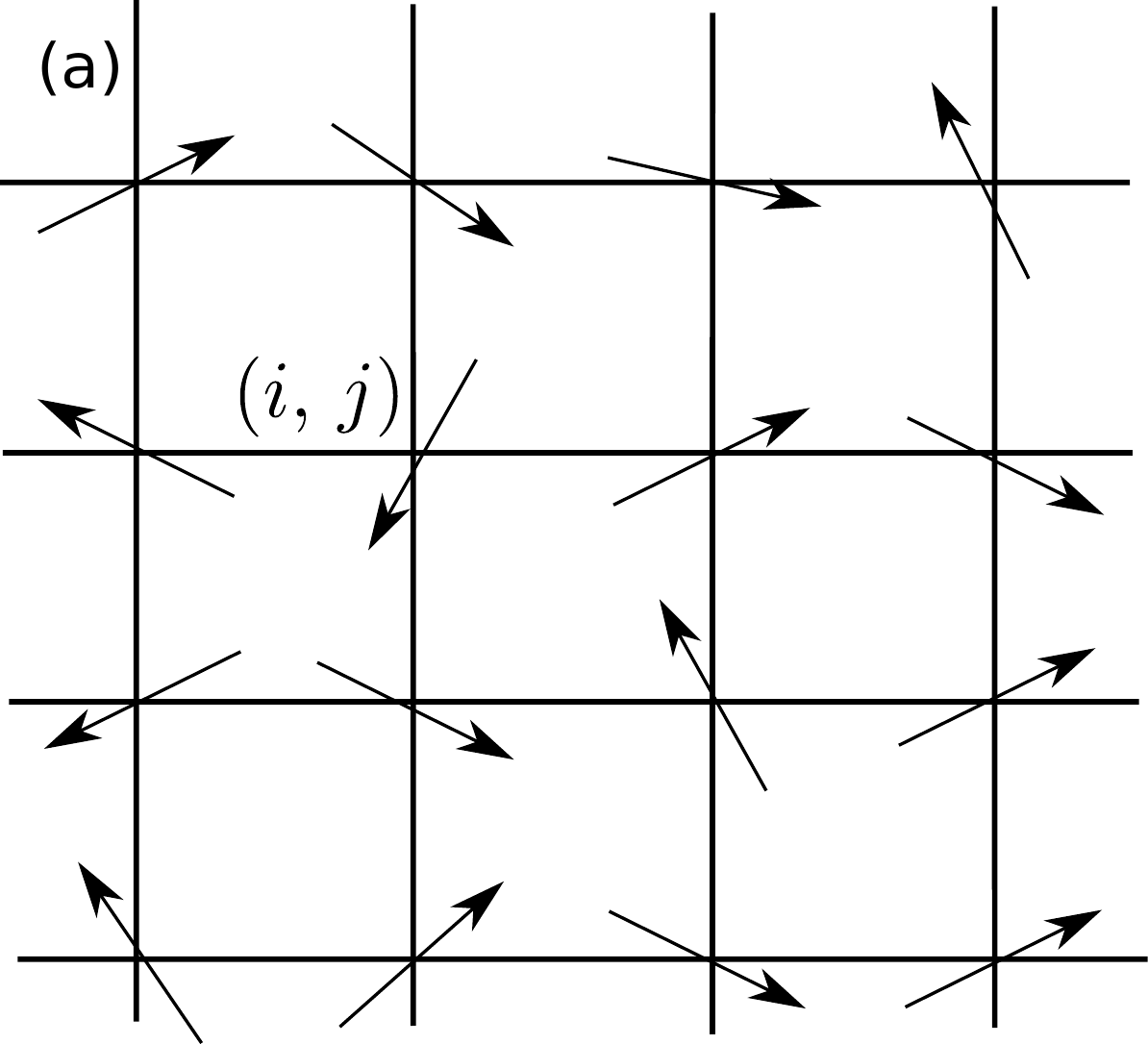}\hfill \includegraphics[width=0.55\columnwidth]{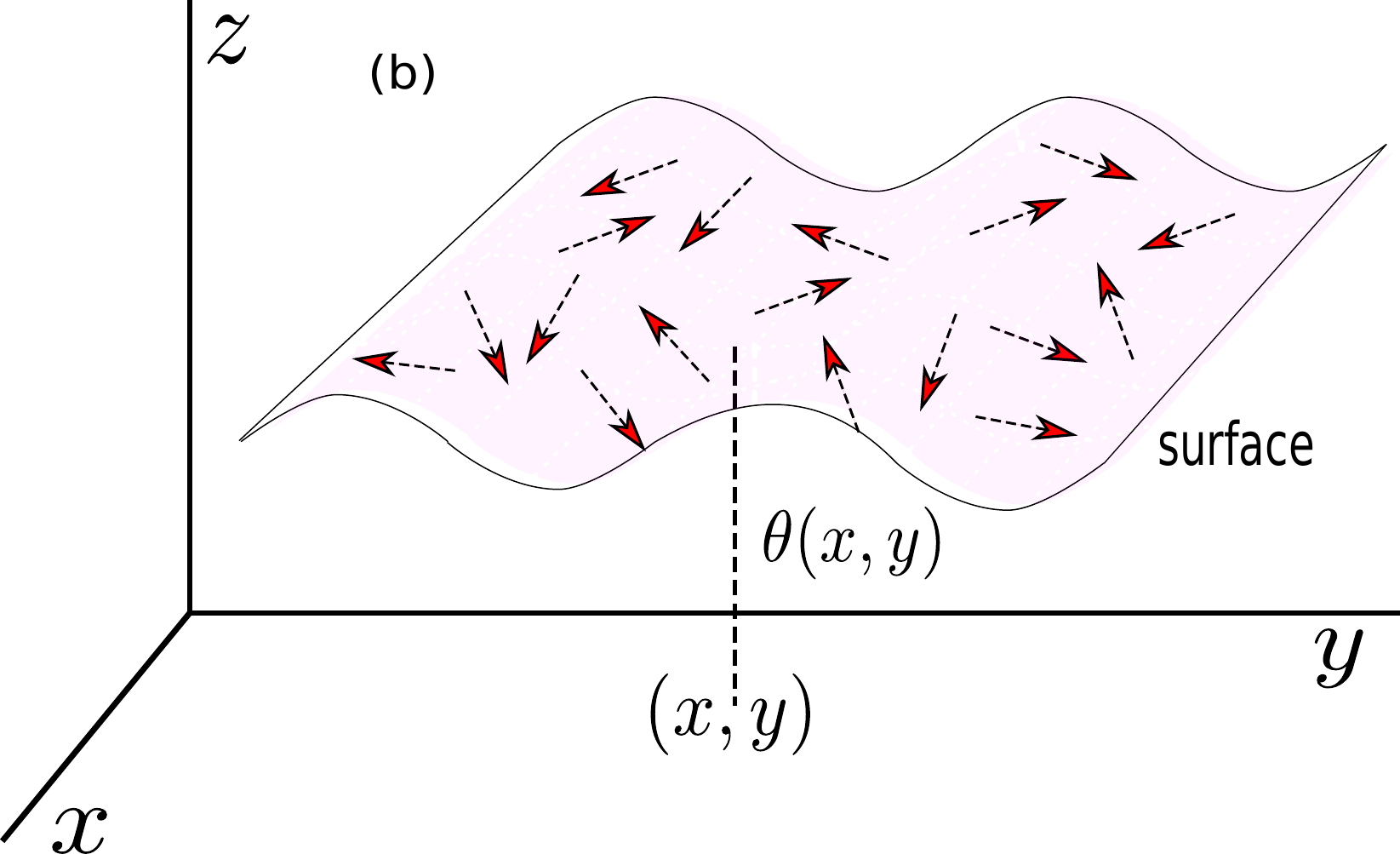}
 \caption{(a) 2D NRB XY model on a square lattice. (b)  An active surface in the Monge gauge coupled with quenched random tangential velocities represented by the arrows.}\label{model}
\end{figure}

The hydrodynamic equation for $\theta({\bf x},t)$, the local phase in the NRB XY model on a substrate and local surface height in the Monge gauge~\cite{nelson-book} is
\begin{eqnarray}
 \frac{\partial\theta}{\partial t} = \kappa{\nabla}^2\theta +\lambda{{\bf F}}\cdot {\boldsymbol\nabla}\theta + \bar\eta,\label{basic-hydro}
\end{eqnarray}
where $\kappa>0$ is the spin stiffness or surface tension, random quenched disordered vector $\bf F$ has its origin in the random nonreciprocal bonds in NRB XY and quenched advection for the surface, $\lambda$ a coupling constant of arbitrary sign. In the absence of any conservation laws or other relevant dynamical variables, $\theta$ is the only slow variable to consider. 
We now derive \eqref{basic-hydro} for the NRB XY model on a substrate, valid  at large spatial and temporal scales. We consider an isotropic, homogeneous frictional medium ensuring
no momentum conservation. In this case, $\theta$ is the only hydrodynamic variable in the problem. We now outline a derivation for \eqref{basic-hydro}, starting from a microscopic model of XY spins grafted rigidly on a 2D square lattice; see Fig.~\ref{model}(a). 
For a spin of fixed length at $(i,j)$ represented by $(\cos \theta_{ij},\sin \theta_{ij})$, the general equation of motion for the angle $\theta_{ij}$, measured with respect to an arbitrary reference line, has the form 
\begin{eqnarray}
 \frac{\partial\theta_{ij}}{\partial t} &&= -\kappa \sum_{\langle i',j'\rangle}\sin (\theta_{ij}-\theta_{i'j'}) \nonumber + \sum_{\langle m\rangle}\tilde F_{i,m}^x\sin (\theta_{ij}-\theta_{mj}) \nonumber \\&&+ \sum_{\langle n\rangle}\tilde F_{j,n}^y\sin(\theta_{ij}-\theta_{in}) + \eta_{ij}. \label{Eq1-basic}
\end{eqnarray}
Here, a subscript $\langle...\rangle$ refers to sum over nearest neighbor sites, $\kappa>0$ is a spin stiffness, $\tilde F^x_{i,m}$ and  $\tilde F^y_{j,n}$ are quenched disordered ``bond vectors'', which link, respectively, the sites $(i,j)$  and $(m,j)$ along the $x$-direction and the sites $(i,j)$ and $(i,n)$ along the $y$-direction.

Now define 
$ \tilde F_{p,q}^{aN} = 
 - \tilde F_{q,p}^{aN}$,
for {\em nonreciprocal} links; $a=x,y$. Next, we replace a bond by a `dual site' by using the convention
$ \tilde F_{m,m+1}^{aN} = -{F}_{m+\frac{1}{2}}^{aN}$.
Now, taking continuum limit,  neglecting higher order and sub-leading terms in the spirit of hydrodynamics, and demanding rotational invariance in 2D (equivalent to isotropicity) we get \eqref{basic-hydro}, 
where for ease of notations, we set $-2 {\bf F}^N\equiv {\bf F}$, and have introduced a coupling constant $\lambda$ for the convenience of the subsequent calculations.  See Appendix for a more detailed derivation of (\ref{basic-hydro}) from (\ref{Eq1-basic}). 

We now show that \eqref{basic-hydro} is also the hydrodynamic equation for an active surface advected by  quenched tangential velocities. Consider a surface whose points in the embedding three-dimensional (3D) space with a position vector ${\bf R} (\vec u)$, where ${\vec u}\equiv (u^1,u^2)$ is an intrinsic coordinate on the membrane; ${\bf R}=(x^1,x^2,x^3)$ is tangentially advected by a 2D velocity field $\vec F=(F^1,F^2)$ defined on the surface. Then the time-evolution of  $\bf R$ follows
\begin{equation}
 \frac{\partial {\bf R}}{\partial t}=\lambda F^a\nabla_a {\bf R} + \kappa \Delta_g {\bf R} + {\boldsymbol {\bar\eta}},
\end{equation}
where $\kappa$ is a surface tension, $\Delta_g {\bf R} = \frac{1}{\sqrt g}\partial_a (\sqrt g g^{ab}\partial_b {\bf R})$, $g_{ab}=\partial_a {\bf R}\cdot \partial_b {\bf R}$ is the induced metric; $a,b=(u^1,u^2)$ are the intrinsic coordinates on the surface. In the Monge gauge~\cite{nelson-book}, appropriate for a nearly flat patch, which is our interest, $a=(x,y), {\bf R}=(x,y,h(x,y)), F^a=(F^x,F^y)$; see Fig.~\ref{model}(b). Then the equation for $h(x,y)$, to the leading order in small slope approximations is (\ref{basic-hydro}).

Equation~\eqref{basic-hydro} models a nonequilibrium or an ``active'' system, since the $\lambda-$ term cannot be obtained from a free energy functional. In fact, \eqref{basic-hydro} implies that for a single realization of $\bf F$ 
\begin{equation}
 \bigg\langle \frac{\partial\theta ({\bf x},t)}{\partial t}\bigg\rangle_F= \bigg\langle \frac{\partial\theta ({\bf k}=0,t)}{\partial t}\bigg\rangle_F=-\lambda\bigg\langle \sum_{\bf q} {\bf F(q)}\cdot i{\bf q} \theta ({\bf -q},t)\bigg\rangle 
\end{equation}
is generally {\em non-zero}, i.e., for a 2D NRB XY model, there is a net rotation of the spins with an angular speed $\omega_0 = \langle \frac{\partial\theta ({\bf k}=0,t)}{\partial t}\rangle_F$, or for a surface, it moves with a drift velocity $v_0 = \langle \frac{\partial\theta ({\bf k}=0,t)}{\partial t}\rangle_F$  due to the active coupling with the quenched tangent vector field. Such clockwise-anticlockwise symmetry breaking global rotation, or up-down symmetry breaking drift are hallmarks of nonequilibrium effects. Here, $\langle...\rangle_F$ implies  
spatial averages for a given realization of $\bf F$. For a constant $\bf F$, the $\lambda$-term in (\ref{basic-hydro}) can be removed by Galilean boost, but not when $\bf F$ is random quenched (see also Ref.~\cite{astik}).

The random quenched disorder field $\bf F$ is assumed to be a zero-mean, Gaussian-distributed with a variance in the Fourier space
\begin{eqnarray}
 \langle F_i({\bf k},\omega) F_j({\bf k}',\omega')\rangle &&= 2 \bigg[D_L Q_{ij}({\bf k}) + D_T P_{ij}({\bf k})\bigg]\nonumber \\ &&\times\delta ({\bf k+k}')\delta(\omega+\omega')\delta(\omega), \label{f-corr}
\end{eqnarray}
where $Q_{ij}({\bf k})=k_ik_j/k^2$ and $P_{ij}({\bf k})=\delta_{ij}-k_ik_j/k^2$ are respectively the longitudinal and transverse projection operators. Both $D_T$ and $D_L$ are positive numbers, a requirement to ensure positivity of the variance of $\bf F(\bf k)$ at {\em all} $\bf k$. The annealed noise $\eta({\bf x},t)$ is zero-mean, Gaussian-distributed with a variance in the Fourier space
\begin{equation}
 \langle\bar\eta({\bf k}),\omega)\bar\eta({\bf k}',\omega')\rangle = 2\bar D \delta ({\bf k+k}')\delta (\omega+\omega').
\end{equation}
Thus, both $\bf F({\bf k})$ and $\bar\eta ({\bf k},\omega)$ are short-ranged. 

In the linearized approximation, i.e., with $\lambda=0$, the equal-time correlations of $\theta({\bf k},t)$ is given by
\begin{equation}
 \langle |\theta({\bf k},t)|^2\rangle_0 = \frac{\bar D}{\kappa k^2},\label{linear-corr}
\end{equation}
where the subscript ``0'' refers to 
a linear theory result.

To go beyond the linear theory, we perform the perturbative dynamic Renormalization Group (hereafter
``RG'') on Eq.~\eqref{basic-hydro} to one-loop order~\cite{stanley,forster,uwe-book}. As usual, fluctuation corrections to the model parameters are represented by the one-loop Feynman graphs
(see Appendix). There are diverging one-loop corrections to $\kappa,\bar D, \lambda$, but not to $D_L,D_T$. Dimensional analysis gives a dimensionless effective coupling constant $g$ and a dimensionless ratio $\mu$:
\begin{equation}
 g\equiv \frac{\lambda^2 D_L}{\kappa^2} \frac{1}{2\pi},\;\mu\equiv \frac{D_T}{D_L}.\label{dim-ratio} 
\end{equation}
Both $g$ and $\mu$ are positive definite. Since $D_L,D_T$ do not renormalize, $\mu$ does not renormalize either. We following the standard steps of RG, which are similar to that for, e.g., the Kardar-Parisi-Zhang equation~\cite{stanley,forster}, with the caveat that $\bf F$ here is time-independent.  In the present model, $\kappa,\bar D,\lambda$ receive RG-relevant corrections to their bare values, all of which diverge logarithimcally in 2D. 
The differential RG recursion relations for $\kappa,\bar D,\lambda$ are 
\begin{eqnarray}
 \frac{d\kappa}{d\ell}&=& \kappa[z-2+ g (\mu-1)],\label{kappa-flow}\\
 \frac{d\bar D}{d\ell}&=& \bar D[z-2-2\chi+2 g],\label{D-flow},\\
 \frac{d\lambda}{d\ell}&=& \lambda[z+\chi_F-1- g].\label{lambda-flow}
\end{eqnarray}
Here, $z,\chi$ are the dynamic and roughness exponents of $\theta$, which describe the spatio-temporal scaling of the correlations of $\theta$~\cite{stanley}, $\chi_F=-1$ gives the spatial scaling of $\bf F$, which can be calculated from (\ref{f-corr}) above. In the linear theory, $z=2,\chi=0$ in 2D. See Appendix for some intermediate steps, which are standard. Lastly, 
$\ell$ is a logarithm of a dimensionless length scale.
The perturbation theory in the present problem is constructed by expanding
in powers of $g$, as can be seen from the
RG recursion relations \eqref{kappa-flow}-\eqref{lambda-flow}. The RG flow equation for $g$ can be obtained from \eqref{kappa-flow}-\eqref{lambda-flow}, giving 
\begin{equation}
 \frac{dg}{d\ell}= -2g^2 \mu.\label{g-flow}
\end{equation}
From the form of (\ref{g-flow}), it is clear that $dg/d\ell<0$ for all $\mu$.  This means starting from an RG ``initial condition'' $g_0=g(\ell=0)$, $g(\ell)$ continues to decay, ultimately reaching zero as $\ell\rightarrow\infty$. This clear from the solution of (\ref{g-flow}), which has a solution 
\begin{equation}
 g(\ell)=\frac{1}{2\mu\ell} \label{large-l-g}
\end{equation}
for large $\ell$. Even though $g=0$ is the only fixed point of (\ref{g-flow}), which is stable, the approach of $g(\ell)$ to 0 is so slow that all of $\kappa,\bar D,\lambda$ receive infinite renormalization, which can be calculated from \eqref{kappa-flow}-\eqref{lambda-flow}. It is convenient to first set $z=2,\chi=0$, the linear theory values, together with $\chi_F=1$ and solve \eqref{kappa-flow}-\eqref{lambda-flow} by using \eqref{large-l-g}. We get renormalized, scale-dependent $\kappa(k),\bar D (k),\lambda (k)$ in the hydrodynamic limit:
\begin{eqnarray}
 \kappa(k)&=&\kappa_0\bigg[\ln\bigg(\frac{\Lambda}{k}\bigg)\bigg]^{\eta_\kappa},\;\eta_\kappa\equiv \frac{\mu-1}{2\mu},\label{kappa-l}\\
 \bar D(\ell)&=& \bar D_0\bigg[\ln\bigg(\frac{\Lambda}{k}\bigg)\bigg]^{\eta_D},\;\eta_D\equiv \frac{1}{\mu},\label{D-l}\\
 \lambda(\ell)&=& \lambda_0\bigg[\ln\bigg(\frac{\Lambda}{k}\bigg)\bigg]^{\eta_\lambda}.\;\eta_\lambda\equiv -\frac{1}{2\mu}.\label{lambda-l}
\end{eqnarray}
Here, $\kappa_0,\bar D_0,\lambda_0$ are the microscopic or unrenormalized values of the respective parameters. Clearly, $\eta_\kappa$ can be of either sign, $\eta_D>0$ and $\eta_\lambda<0$ necessarily. This means in the long wavelength limit, the renormalized version of (\ref{basic-hydro}), written in terms of the renormalized parameters, is effectively linearized or noninteracting! This allows us to calculate the renormalized correlation function for $\theta({\bf k},t)$:
\begin{equation}
 \langle |\theta({\bf k},t)|^2\rangle = \frac{\bar D(k)}{\kappa(k) k^2}=\frac{D_0}{\kappa_0 k^2} \bigg[\ln \bigg(\frac{\Lambda}{k}\bigg)\bigg]^{\eta_D-\eta_\kappa},\label{renorm-corr}
\end{equation}
see also Refs~\cite{pelcovich,sm-ising-elastic1,sm-ising-elastic2,diffxy1,diffxy2,debayan1}.
The real space phase fluctuations of $\theta({\bf x},t)$ can be calculated by
Fourier transforming~\eqref{renorm-corr} back to real space, giving \eqref{basic}, where
\begin{equation}
 \eta=\eta_D - \eta_\kappa +1.\label{eta}
\end{equation}
Exponent $\eta$ varies continuously with $\mu$. 
Since $\eta_D - \eta_\kappa=1/\mu-(\mu-1)/(2\mu)$ can be positive or negative, $\eta$ can be more or less 1. Using \eqref{kappa-l}, \eqref{D-l} and \eqref{eta}, $\eta_\text{min}=1/2$ gives the minimum of $\eta$, when $\mu\rightarrow\infty$, corresponding to $\bf  F$ being fully solenoidal. In this case, 
\begin{equation}
 \langle \theta({\bf x},t)^2\rangle = \frac{D_0}{2\pi\kappa_0}\sqrt{\ln \bigg(\frac{L}{a_0}\bigg)},
\end{equation}
which though rises with $L$, ultimately diverging as $L\rightarrow\infty$ ruling out true long range order, it does so {\em infinitely slowly} relative to QLRO. Thus one has only quasi-phase order here, although it can be infinitely stronger than conventional QLRO in 2D XY model or in an EW surface. 
In the opposite limit, for $\mu\rightarrow 0$ for a fully irrotational $\bf F$, $\eta\rightarrow \infty$, giving large $\theta$-fluctuations, {\em infinitely} larger than those in QLRO. Therefore, $1/2\leq \eta $ with {\em no upper bound}.

The relaxation time-scale $\tau(k)= [\kappa(k)k^2]^{-1}$ of $\theta$-fluctuations in the long wavelength limit display novel behavior. Since $\eta_\kappa$ can be positive or negative, $\tau(k)$ can be smaller or bigger than usual diffusive dynamics, giving faster or slower than diffusive dynamics: For $\mu\rightarrow\infty$ (fully irrotational), $\eta_\kappa\rightarrow 1/2$ is its maximum value; as $\mu\rightarrow 0$, $\eta_\kappa \rightarrow -\infty$, giving $\eta_\kappa\geq 1/2$ with {\em no lower bound}. Interestingly, as $\mu\rightarrow 0$, $\eta\rightarrow\infty,\,\eta_\kappa\rightarrow-\infty$, i.e., very little order with very slow dynamics, which is reminiscent of spin-glass like behavior~\cite{spin-glass}. See Fig.~\ref{phase}(b).

Overall then, (i) $\mu>3$ (i.e., $D_T>3D_L$), $1/2\leq \eta<1$ and $\eta_\kappa>0$, giving strongly ordered states with faster relaxation, (ii) $1<\mu<3$ (i.e., $D_L<D_T<3D_L$), $\eta>1$ and $\eta_\kappa>0$, giving weakly ordered states with faster relaxation,(iii) $0\leq \mu<1$ (i.e., $D_T<D_L$), $\eta>1$ and $\eta_\kappa<0$, giving weakly ordered states with slower relaxation. At $\mu=3$, $\eta=1,\eta_\kappa>0$ giving QLRO with faster relaxation, and at $\mu=1$, $\eta>1,\eta_\kappa=0$, giving weakly phase-ordered states with diffusive dynamics. Thus at no value of $\mu$, the renormalized spatio-temporal scaling of $\theta$ in the long wavelength limit is identical  to the linear theory results. These results are summarized in the phase diagram drawn in the $D_L-D_T$ plane in Fig.~\ref{phase}(a). 

We now argue that although the above results were obtained from a one-loop perturbation theory, they are actually {\em exact} in the long wavelength limit. First of all, the nonuniversality appears due to $\mu$ being marginal at the one-loop order. This stems from the nonrenormalization of $D_L,D_T$. From the structure of the theory, the correlator of $\bf F$ in \eqref{f-corr} and hence $\mu$ must remain unrenormalized to any order in the perturbation theory. Thus the nonuniversal aspect is robust. Secondly, any higher loop contributions to $\kappa,\bar D,\lambda$ should be proportional to powers of $g$ more than 1. These higher-loop corrections can potentially give rise to additional contributions to \eqref{g-flow} at higher powers in $1/\ell$. These in turn can generate {\em finite} corrections to $\kappa,\bar D,\lambda$, in the long wavelength limit, which can be neglected in the spirit of RG. This makes the results on scaling from the one-loop perturbation theory exact in the long wavelength limit.

We now argue that vortex unbinding transitions  take place for $\eta>1$ in the NRB XY model, destroying the ordered states. To study it systematically, it is necessary to include vortices
in our description of the system. 
In the absence of any systematic theory to handle vortices in active XY models, we use a workaround. First of all given that  $\beta(\ell)\equiv\kappa(\ell)/\bar D(\ell)$ plays the role of inverse reduced temperature $\kappa/(k_BT)$ in the equilibrium 2D XY model~\cite{chaikin}, reduced temperature $\beta(\ell)^{-1}$ vanishes (diverges) for $\eta_\kappa>(<)\eta_D$, i.e., $\eta< (>)1$ in the long wavelength limit. This should imply states with {\em no} vortices, as in the 2D XY model in its low temperature QLRO phase (states full of vortices, as in the 2D XY model in its high temperature short range ordered phase). This suggests that the weakly phase-ordered states with $\eta>1$ are destroyed by vortex proliferation in the NRB XY model.
This argument is more formally stated by considering the recursion relation of the vortex fugacity $y$ in the equilibrium 2D XY model along with our recursion relations (\ref{kappa-flow})-(\ref{lambda-flow}) above:
\begin{equation}
 \frac{dy}{d\ell}=\left[2- \beta(\ell)\right]=\left[2-\beta_0\ell^{\eta_\kappa-\eta_D}\right];
\end{equation}
$\beta_0\equiv \beta(\ell=0)$.
In the large $\ell$-limit, $\beta(\ell)\rightarrow 0 (\infty)$ for  $\eta<1(>1)$, the vortices are {\em always} unbounded for $\eta>1$, destroying the ordered states.
Defining a length scale $\tilde \ell$ at which $dy/d\ell=0$, we get $\tilde \ell= (2/\beta_0)^{1/{\eta_\kappa-\eta_D}}$, giving a correlation length $\xi=a_0\exp[ (2/\beta_0)^{6/|\delta|}],\,\delta\equiv \mu-3$.

In summary, we have shown that 2D nonreciprocal random bond XY model and active surfaces with quenched tangential velocities are described by the same hydrodynamic theory. This theory predicts that these models can admit order better than QLRO of 2D XY model, giving sub-logarithmic roughness with nonuniversal exponents. The surface can also show super-logarithmic roughness, again with nonuniversal exponents, which should be absent in the NRB XY model due to vortex proliferations.



{\em Acknowledgment:-} S.M. and A.B. thank Alexander von Humboldt Stiftung (Germany)
for partial financial support through their research group linkage programme (2024). S.M. thanks ANRF (India) for partial financial
support through the ARG (MATRICS) programme (file no.: ANRF/ARGM/2025/000748/TS).  A.B. thanks ANRF (India) for partial financial
support through the ARG (MATRICS) programme (file no.:
ANRF/ARGM/2025/000461/TS).

\newpage

\begin{widetext}

\appendix 

\section{Derivation of the hydrodynamic equation for the nonreciprocal random bond XY model}

We consider a  2D square lattice; see Fig.~2(a) of the main text. Now assign an XY spin of fixed length unity at each lattice point 
$(i,j)$. We represent the spin at $(i,j)$ by $(\cos \theta_{ij},\sin \theta_{ij})$, i.e., by a phase angle $\theta_{ij}$. Next, we define quenched ``bond vector'' between two bonds. Along the $x$  direction such a bond vector is $\tilde F^x_{i,m}$, i.e., $\tilde F^x_{i,m}$ is the bond vector between $(i,j)$  and $(m,j)$ sites. Similarly, along $y$ direction a bond vector is $\tilde F^y_{j,n}$ connecting the sites $(i,j)$ and $(i,n)$. Each spin has nearest neighbour interactions only, appropriately generalized for bond disorders. The dynamical equation for $\theta_{ij}$ is
\begin{eqnarray}
 \frac{\partial\theta_{ij}}{\partial t} = -\kappa \sum_{\langle i',j'\rangle}\sin (\theta_{ij}-\theta_{i'j'}) + \sum_{\langle m\rangle}\tilde F_{i,m}^x\sin (\theta_{ij}-\theta_{mj}) + \sum_{\langle n\rangle}\tilde F_{j,n}^y\sin(\theta_{ij}-\theta_{in}) + \eta_{ij}. \label{Eq1-basic-SM}
\end{eqnarray}
Considering nearest neighbour interactions,  we obtain 
\begin{eqnarray}
 \frac{\partial\theta_{ij}}{\partial t} &=& -\kappa [\sin (\theta_{ij}-\theta_{i-1j}) + \sin (\theta_{ij}-\theta_{i+1j})+ \sin (\theta_{ij}-\theta_{ij-1})+\sin (\theta_{ij}-\theta_{ij+1})]  + \tilde F_{i,i-1}^x\sin (\theta_{ij}-\theta_{i-1j}) \nonumber \\ 
 &&\;\;\;\;\;\; + \tilde F_{i,i+1}^x\sin (\theta_{ij}-\theta_{i+1j})+ \tilde F_{j,j-1}^y\sin(\theta_{ij}-\theta_{ij-1}) + \tilde F_{j,j+1}^y\sin(\theta_{ij}-\theta_{ij+1}) + \eta_{ij}. \label{Eq1}
\end{eqnarray}
We define 
\begin{equation}
\tilde F_{p,q}^a = \tilde F_{p,q}^{aR} + \tilde F_{p,q}^{aN}~~~~~~~~~a=x,y, 
\end{equation}
such that 
\begin{equation}
 \tilde F_{p,q}^{aR} = \tilde F_{q,p}^{aR}~~\text{(reciprocal)}, ~~~~~\tilde F_{p,q}^{aN} = - \tilde F_{q,p}^{aN}~~\text{(nonreciprocal)}
\end{equation}
Now we replace a bond by a ``dual site'' by using the convention
\begin{equation}
 \tilde F_{m,m+1}^{aR} = {F}_{m+\frac{1}{2}}^{aR}~~\text{and}~~\tilde F_{m,m+1}^{aN} = -{F}_{m+\frac{1}{2}}^{aN}
\end{equation}
Using this convention we get
\begin{eqnarray}
 \tilde F_{i,i+1}^{x} &=& \tilde F_{i,i+1}^{xR} + \tilde F_{i,i+1}^{xN} \nonumber\\
 &=& {F}_{i+\frac{1}{2}}^{xR} + {F}_{i+\frac{1}{2}}^{xN}
\end{eqnarray}
and
\begin{eqnarray}
 \tilde F_{i,i-1}^{x} &=& \tilde F_{i,i-1}^{xR} + \tilde F_{i,i-1}^{xN} \nonumber\\
 &=& F_{i-1,i}^{xR} - F_{i-1,i}^{xN}\nonumber\\
 &=& {F}_{i-\frac{1}{2}}^{xR} - {F}_{i-\frac{1}{2}}^{xN}
\end{eqnarray}
Similarly we get 
\begin{eqnarray}
\tilde F_{j,j+1}^{y} = {F}_{j+\frac{1}{2}}^{yR} + {F}_{j+\frac{1}{2}}^{yN}~~~\text{and}~~~\tilde F_{j,j-1}^{y} = {F}_{j-\frac{1}{2}}^{yR} - {F}_{j-\frac{1}{2}}^{yN}.
\end{eqnarray}
Therefore we get
\begin{eqnarray}
 &&\tilde F_{i,i-1}^x\sin (\theta_{ij}-\theta_{i-1j})+\tilde F_{i,i+1}^x\sin (\theta_{ij}-\theta_{i+1j}) \nonumber \\
 &=&\Big{(}{F}_{i-\frac{1}{2}}^{xR} - {F}_{i-\frac{1}{2}}^{xN}\Big{)}\sin (\theta_{ij}-\theta_{i-1j})+\Big{(}{F}_{i+\frac{1}{2}}^{xR} + {F}_{i+\frac{1}{2}}^{xN}\Big{)}\sin (\theta_{ij}-\theta_{i+1j}) \nonumber\\
 &\approx&\Big{(}{F}_{i-\frac{1}{2}}^{xR} - {F}_{i-\frac{1}{2}}^{xN}\Big{)}(\theta_{ij}-\theta_{i-1j})+\Big{(}{F}_{i+\frac{1}{2}}^{xR} + {F}_{i+\frac{1}{2}}^{xN}\Big{)}(\theta_{ij}-\theta_{i+1j}), 
\end{eqnarray}
and similarly in the $y$-direction.
We take a continuum limit and denote space point by ${\bf r}={\bf r}(x,y)$. Considering continuum limit and neglecting higher order and sub-leading terms we get 
\begin{eqnarray}
\tilde F_{i,i-1}^x\sin (\theta_{ij}-\theta_{i-1j})+\tilde F_{i,i+1}^x\sin (\theta_{ij}-\theta_{i+1j})
&\approx& -2{F}^{xN}{\partial}_x \theta - {\partial}_x {F}^{xR}{\partial}_x \theta - {F}^{xR}{{\partial}^2_x} \theta
\end{eqnarray}
Similarly we obtain 
\begin{eqnarray}
 \tilde F_{j,j-1}^y\sin(\theta_{ij}-\theta_{ij-1}) + \tilde F_{j,j+1}^y\sin(\theta_{ij}-\theta_{ij+1}) &\approx& -2{F}^{yN}{\partial}_y \theta - {\partial}_y {F}^{yR}{\partial}_y \theta - {F}^{yR}{{\partial}^2_y} \theta.
\end{eqnarray}
The continuum equation for $\theta ({\bf r})$ now reads 
\begin{eqnarray}
 \frac{\partial\theta}{\partial t} &=& \kappa\Big{(}{\partial}^2_x+{\partial}^2_y\Big{)}\theta -2{F}^{xN}{\partial}_x \theta - {\partial}_x {F}^{xR}{\partial}_x \theta - {F}^{xR}{{\partial}^2_x} \theta -2{F}^{yN}{\partial}_y \theta - {\partial}_y {F}^{yR}{\partial}_y \theta - {F}^{yR}{{\partial}^2_y} \theta \nonumber \\
 &=& \kappa\Big{(}{\partial}^2_x+{\partial}^2_y\Big{)}\theta -2\Big{(}{F}^{xN}{\partial}_x \theta + {F}^{yN}{\partial}_y \theta\Big{)} - {\partial}_x {F}^{xR}{\partial}_x \theta - {\partial}_y{F}^{yR}{\partial}_y \theta -{F}^{xR}{{\partial}^2_x} \theta - {F}^{yR}{{\partial}^2_y} \theta + \eta.\label{conti-theta}
\end{eqnarray}
So far we have not considered any  symmetry of the system. In absence of any breakdown of parity,  Eq.~(\ref{conti-theta}) must remain invariant under coordinate inversions $x \rightarrow - x$ and $y \rightarrow - y$. This requires 
\begin{eqnarray}
{F}^{xN}(-x) \rightarrow -{F}^{xN}(x)~~\text{and}~~{F}^{yN}(-y) \rightarrow -{F}^{yN}(y) \nonumber\\
{F}^{xR}(-x) \rightarrow {F}^{xR}(x)~~\text{and}~~{F}^{yR}(-y) \rightarrow {F}^{yR}(y) \nonumber
\end{eqnarray}
Thus ${F}^{N} = ({F}^{xN},{F}^{yN})$ transforms like a vector under coordinate inversion whereas ${F}^{xR},{F}^{yR}$ transform like scalars. 

We now demand the invariance under 2D rotation in the plane (which is isotropic assumption). Let us consider a rotation by an angle $\pi/2$. Under this rotation the coordinates transform as $x \leftrightarrow y$. Invariance under this requires ${F}^{xN} \leftrightarrow {F}^{yN}$ and ${F}^{xR} = {F}^{yR} = \phi$ (say). Therefore Eq.~(\ref{conti-theta}) becomes
\begin{eqnarray}
 \frac{\partial\theta}{\partial t} = \kappa{\nabla}^2\theta - 2{\tilde{\bf F}}^{N}\cdot {\boldsymbol\nabla}\theta - ({\boldsymbol\nabla}\phi)\cdot({\boldsymbol\nabla}\theta) - \phi{\nabla}^2\theta + \bar\eta
\end{eqnarray}
Introducing arbitrary coupling constants we write 
\begin{eqnarray}
 \frac{\partial\theta}{\partial t} = \kappa{\nabla}^2\theta + {\lambda}{{\bf F}}^{N}\cdot{\boldsymbol\nabla}\theta + {\lambda}_2({\boldsymbol\nabla}\phi)\cdot({\boldsymbol\nabla}\theta) + {\lambda}_2\phi{\nabla}^2\theta + \bar\eta. \label{gen-eom}
\end{eqnarray}
The coupling constants for the reciprocal terms are same to reflect the fact that both the terms originate from the same discrete, microscopic terms in the lattice equation.
The above equation has nonreciprocal and reciprocal quenched disordered terms. Assuming both ${\bf F}^N$ and $\phi$ have the same microscopic origin, and hence have the same spatial scaling, the nonreciprocal term is clearly more relevant (in the scaling sense). If we now identify ${\bf F}^N$ with $\bf F$ and ignore the reciprocal terms with $\phi$ (which are subleading to the nonreciprocal terms), then we get Eq.~(3) of the main text.

If the random bond interactions are purely reciprocal then Eq.~(\ref{gen-eom}) becomes
\begin{eqnarray}
 \frac{\partial\theta}{\partial t} = \kappa{\nabla}^2\theta  + {\lambda}_2{\boldsymbol\nabla}(\phi{\boldsymbol\nabla}\theta) + \bar\eta. 
\end{eqnarray}
This is an equilibrium model with a relaxational dynamics of a free energy functional
\begin{equation}
 {\cal  F} = \int d^{d}{\bf r}~\Big[\frac{\kappa}{2}({\boldsymbol\nabla}\theta)^2 + \frac{\lambda_2}{2}\phi({\boldsymbol\nabla}\theta)^2\Big].
\end{equation}
We can obtain the equation of motion as
\begin{equation}
\frac{\partial\theta}{\partial t} = - \frac{\delta {\cal F}}{\delta \theta} + \bar\eta. 
\end{equation}

\section{Details of the perturbative renormalization group calculations}

The perturbation theory is constructed by expanding the solution of the Eq.~(3) in powers of the nonlinear copupling constant $\lambda$~\cite{stanley}. The propagator $G_0(k,\omega)$ and correlator $C_0(k,\omega)$  of $\theta$ in Fourier space in the linear theory ($\lambda=0$) are 
\begin{eqnarray}
 G_0 (k,\omega)&=& \frac{1}{-i\omega + \kappa kq^2},\\
 C_0(k,\omega) &=& \frac{2\bar D}{\omega^2 + \kappa^2 k^4},
\end{eqnarray}
where a subscript ``0'' refers to the expressions being evaluated in the linear theory.
The perturbation theory is conveniently handled by using a path integral formulation in terms of a generating functional~\cite{uwe-book}, in which $\lambda$ appears as an anharmonic coupling constant, by introducing a dynamic conjugate variable $\hat\theta$ and with the identification
\begin{equation}
 G(k,\omega) = \langle \hat \theta(-{\bf k},-\omega)\theta({\bf k},\omega)\rangle.
\end{equation}
An expansion of the generating functional in powers of $\lambda$ has a one-to-one correspondence with the expansion in $\lambda$ in the equation of motion approach~\cite{stanley} to all orders in the perturbation theory,

\begin{figure}[htb]
 \includegraphics[width=0.40\textwidth]{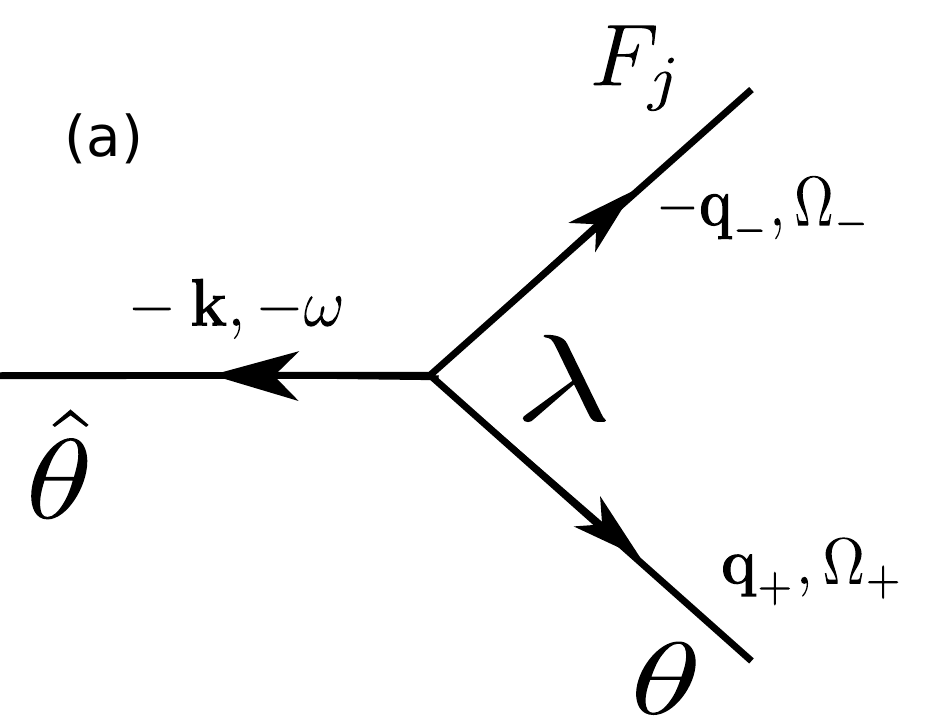}\includegraphics[width=0.60\textwidth]{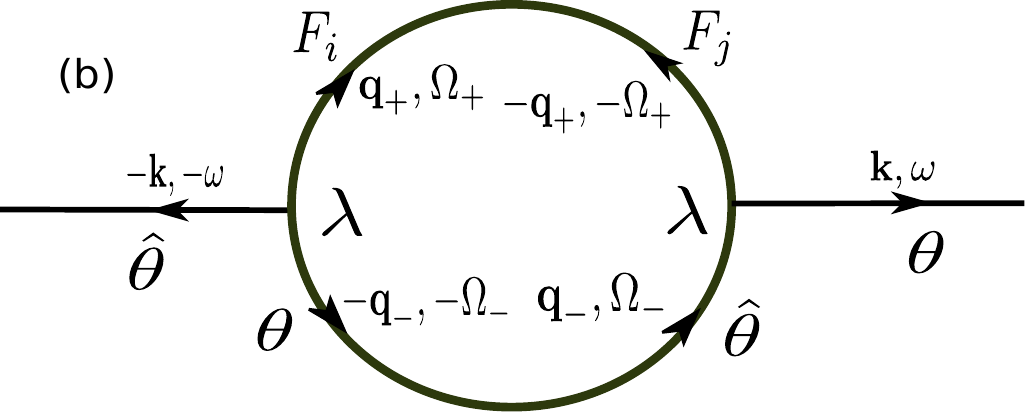}\\
 \includegraphics[width=0.60\textwidth]{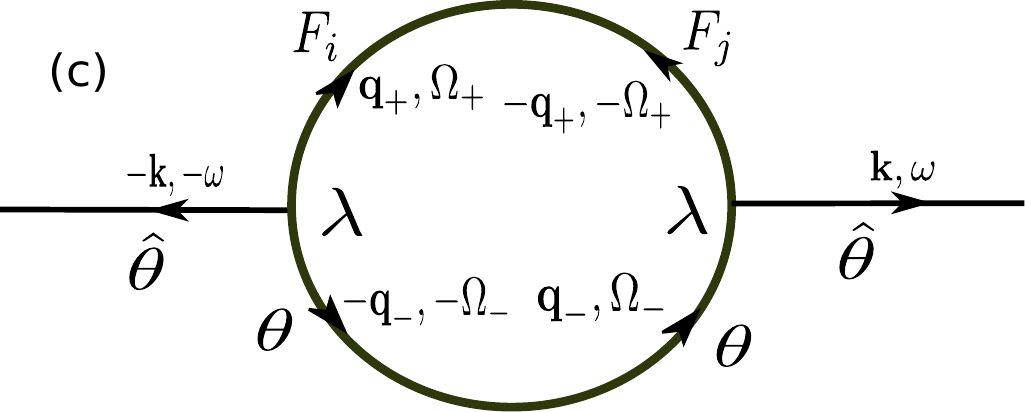}
 \includegraphics[width=0.35\textwidth]{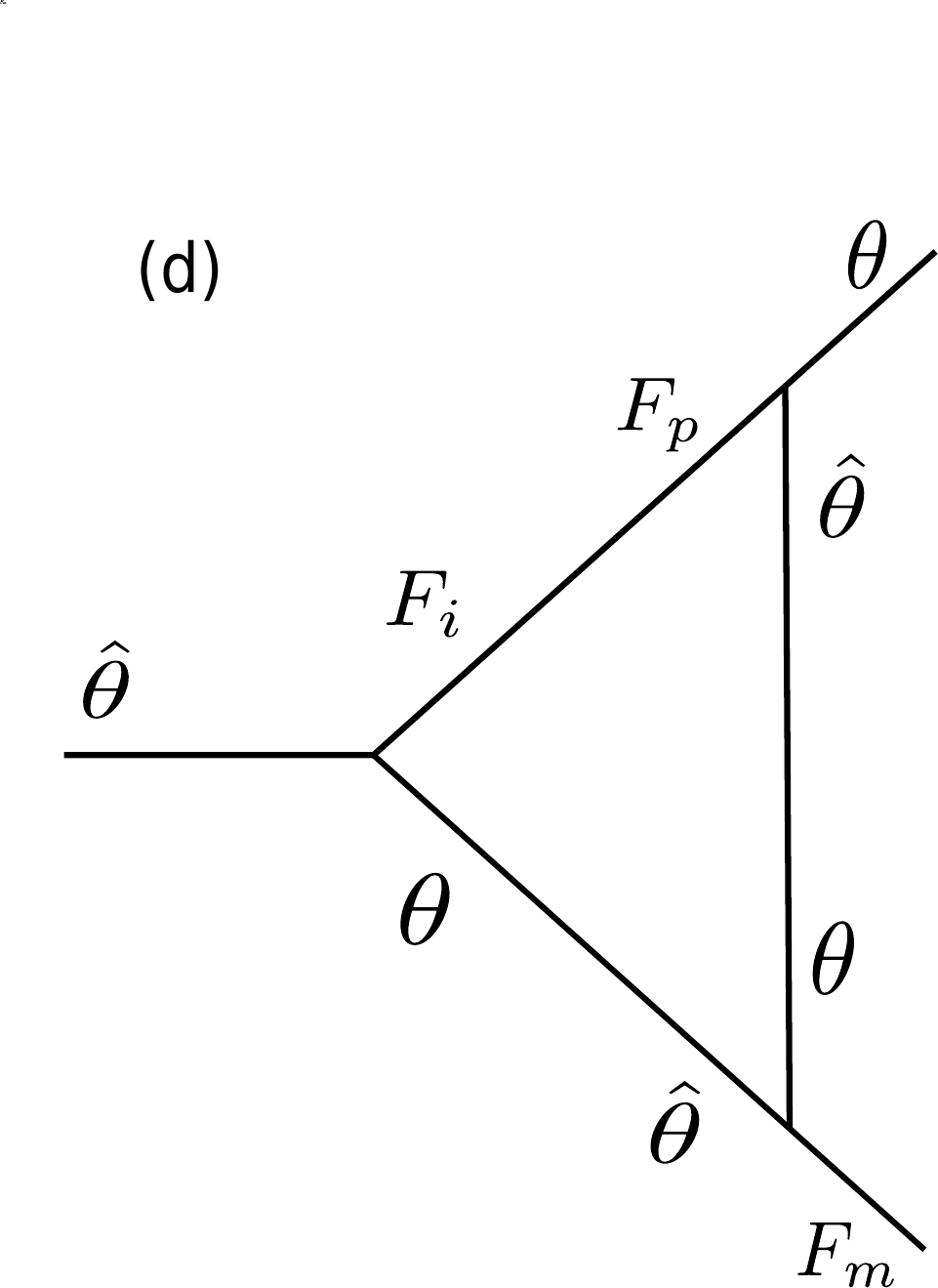}
 \caption{Feynman graphs for (a) Tree-level vertex, (b) one-loop propagator, (c) one-loop correlator, (d) one-loop vertex.}\label{feyn-graphs}
\end{figure}
The tree-level vertx and the one-loop graphs for $\kappa,\bar D,\lambda$ are given in Fig.~\ref{feyn-graphs},
The one-loop corrections to the $\kappa,\bar D,\lambda$ are given by
\begin{eqnarray}
 \Delta \kappa &=&\frac{\lambda^2}{2\pi}\frac{D_T-D_L}{\kappa}\ln b,\\
 \Delta \bar D &=& \frac{2\lambda^2}{2\pi} \frac{D_L}{\kappa^2}\bar D \ln b,\\
 \Delta \lambda &=&-\frac{D_L\lambda^3}{2\pi\kappa^2}\ln b.
\end{eqnarray}

\end{widetext}

\end{document}